\documentclass[12pt,a4paper]{article}

\usepackage[lmargin=2cm,rmargin=2.5cm,tmargin=3cm,bmargin=2cm]{geometry}
\usepackage{enumitem}
\begin{document}
\title{The Bomb, the Hubris, and the Spectre}
\author{Alberto Girlando\thanks{e-mail: alberto.girlando@momag.it; www: albertogirlando@wordpress.com.}\\ \\ \small{Molecular Materials Group (MoMaG), 43124 Parma, Italy}}

\date{}

\maketitle

\begin{abstract}
In this essay I will make a non-exhaustive survey of the role of scientists, politicians, and the military in the development and use of the Bomb in the thirty years following the beginning of the United States Manhattan Project. I will be mainly focusing on facts or documents that have gradually fallen into oblivion, but that in my opinion are important to develop a critical view of the past responsibilities, and to consciously address the present dramatic situation of worldwide wars, that may yield again to the use of the Bomb. 
\end{abstract}

More than 80 years have passed since the Bomb destroyed Hiroshima and Nagasaki, causing  worldwide horror and fear, feelings that eventually led to a gradual dismissal of  nuclear stockpiles by the main nuclear powers, United States and URSS-Russia. In addition, dramatic incidents in nuclear plants induced some countries - Italy and Germany, for instance - to also ban  nuclear power for civil use. Since it is well know that civil plants are the prerequisite, if not the cover-up, for the development of the Bomb, people of my generation started to believe that a nuclear war had become a remote possibility.

But nowadays we have to face a different reality: nuclear power for military use is again considered a realistic outcome of the present chaotic war scenario, with adverse sentiments and fears slowly fading away. Not only the military, but also politicians and opinion makers have started to talk about the legitimate use of “tactic” nuclear bombing, namely, a Bomb of limited power, and directed to military installations only. But the target, military rather than civil (the latter is called a “strategic” objective), is the only difference, as the supposed “limited power” that would be possibly used is actually the same as that of the Hiroshima Bomb, i.e. $\sim$20 kilotons (kT), that at the time led to about 200 000 causalities.

In an attempt to understand how this nuclear war scenario has become
again "acceptable", I started to investigate the facts following Hiroshima, and
found that the basic reasons for the use of the Bomb had never disappeared, but were simply hidden away. The present notes represent my current personal view, emerging from casual reading of scientific biographies, history books and articles connected to the development of the Bomb, now and then assisted by searches in Internet. The covered historical period roughly goes from 1939 to 1968,
i.e. from the first suggestion that an atomic Bomb was possible, until the Treaty on the Non-Proliferation of Nuclear Weapons (NPT). The picture coming out from these reading is rather different from the account I had been used to, with several facts that had faded away from the Occidental narrative of this historical period.   

Of the five countries winners of the Second World War, United States (US), United Kingdom (UK), Soviet Union (URSS), Cina and France, only the US emerged stronger than before the war: the war did not enter US own territory, the industry had become stronger, and, most and foremost, US, and only US, had the Bomb, the ultimate and definitive weapon. Within a few years, US had become the dominant nation in the world, and under those circumstances, the “hubris” - the danger of which ancient Greece had warned the powerful about - naturally took hold: no other country could stand up to the US overwhelming power. 

Of course, to retain such a dominant position, US had to maintain the advantage of the Bomb: nor the allied nor the enemies should be able to develop one. Germany and Japan had been fully defeated, so where were the enemies? Well, the enemy was an old one, the “Spectre  that is haunting Europe” as stated in the previous century by Marx and Engels in the “Manifesto of the Communist Party”.  And at the time Communism was the political regime of Soviet Union, a nation that had become a pro-tempore United States ally thanks to Hitler.

\vskip 0.5 cm

I shall start from the famous letter written by Einstein to US President Roosevelt in August 1939, just a month before the beginning of war in Europe: “Some recent work by Enrico Fermi and Leo Szilard ...led me to expect that the element Uranium may be turned into a new and important source of energy.... In the course of the last four months it has been made probable through the work of Joliot in France, as well as Fermi and Szilard in America, that it may be possible to set up a nuclear chain reaction in a large mass of Uranium.... this new phenomenon would also lead to the construction of bombs...”  (letter by Einstein to Roosevelt, 2 August 1939).

There are two facts worth highlighting in this letter. First, all the three scientist mentioned were Europeans, as quantum mechanics was entirely developed in Europe, and American PhD students like Oppenheimer came to Europe to learn. But tanks to Nazi-fascism, most of them fled abroad, particularly to the United States, that progressively began the new center of science. As a matter of fact Einstein, Fermi and Szilard (the inspirator behind the Einstein's letter) were already there. Second, the first scientist mentioned in the letter, the French Nobel prize Joliot, and his team, was probably the most advanced in the effort to harness the atomic energy, but he did remain in France, also after the Nazi invasion. Of course, he suspended his research on nuclear energy and, drawing on his fame, prestige, and technical skills, actively aided the French "maquis" resistance. He resumed nuclear research after the war, and under his directive France built his first nuclear reactor as early as 1948. Despite of this scientific and - I would add -  moral record, his figure quickly faded away also in France, for reasons I shall present later.

Following the Einstein-Szilard recommendation, Roosevelt set up the Manhattan project. During the first two years, however, while war was breaking out in Europe, the project didn’t received much financial and political support, partly because scientists still knew too little about the physics involved. Parallel efforts were pursued all over Europe, notably in England, URSS, and Germany. But in Germany, as well as in Italy, the most distinguished scientists had left the country, mainly because they were of Jewish origin, or because they were communist or in any case opposing Hitler and Mussolini dictatorship. In Germany there was also the “Deutsche Physik”  movement, that rejected quantum mechanics as being of Jewish origin. Heisenberg was the only Nobel laureate remaining in his country, but he was a pure theoretician and his role, if  any, in the possible attempt to build the Bomb by Germany remained obscure. 

\newpage
At the beginning of the war it was widely believed that obtaining a nuclear bomb was a nearly impossible feat, requiring huge amounts of Uranium. But all that changed when two refugees in England, Frisch and Peierls, realized and calculated that using the rare isotope 135 of Uranium rather than the common 138 isotope, it was indeed possible to build a bomb. Peierls  personally typed the resulting report, now known as the Frisch-Peierls memorandum, drafted in March 1940. Through personal contacts among scientists, also involved in other wartime projects, such as Mark Oliphant, involved in radar research, the breaking news reached Ernest Lawrence in the US, and through him Roosevelt’s scientific advisor, Vannevar Bush. In October 1941 Bush convinced Roosevelt to fund and expand the original Manhattan Project (e.g., by establishing the Los Alamos site, and so on).  Two months later, on December 1941, Japan declared war on the United States, followed immediately by Germany and Italy. The United States then officially entered the war, and since Hitler had attacked the Soviet Union in June 1941, the Soviet Union became an indirect ally of the United States and UK - with great mistrust on both sides.

After the Roosevelt approval, Bush appointed General Grooves, an engineer by formation, as the overall coordinator of the Manhattan project, now officially tasked with building the Bomb (1942). In turn, Grooves nominated Robert Oppenheimer as scientific director. With the military's involvement, overall secrecy was strictly adopted, also towards the English scientists that had been instrumental in promoting the atomic Bomb program in the US. It was not until a year later, in 1943, that Churchill convinced Roosevelt to accept collaboration of UK scientists.  Churchill’s argument was that only by working together could they win the race against Germany or the USSR in building the Bomb. Thus, a British delegation, led by Chadwick and including Rudolf Peierls, traveled to the United States to collaborate with American scientists. But Grooves, by compartmentalizing the research staff, did not allow the British team to have a complete picture of the Bomb project.

As a consequence of the Smith Act (1940), all individuals involved in the Manhattan Project were required to undergo screening to verify their lack of affiliation with “organizations that advocated the violent overthrow of the Government.” That rule could be applied both to fascists and communists, but it was used mainly against communists. So all the scientists working on the Manhattan project had to be “cleared” for communist ideas or sympathies. But ties, connections or sympathies among scientists in the international team were unavoidable. Certainly all scientist working at the Manhattan project were opposed to Hitler: Nazism had entrenched the idea of world domination through violence on “inferior” populations. Not so Communism. Oppenheimer himself was sympathetic of communist ideas, and at the time relatively little was known about Stalin’s regime. 

So when, around March 1944, Grooves stated during informal discussions with the leaders of the Manhattan Project that the true purpose of building the Bomb was to subdue the Soviets, many scientists were appalled. One of them, Joseph Rotblat, a Polish refugee of the English team, decided to quit immediately. In the end, he was allowed to do so, even though US intelligence agencies had tried to fabricate a dossier - which turned out to be blatantly false - accusing him of espionage for the Soviets. Rotblat was the only one to leave; the others stayed although it was clear that after Stalingrad (February 1943) and Italy armistice (September 1943) Germany would be likely defeated. It was equally clear, however, that the Manhattan Project would not be dismantled in the event of German defeat before the Bomb was ready.

In fact two years later, in 1945, the war was nearing its end, and in February  Roosevelt, Churchill and Stalin met in Yalta to discuss the post-war world reorganization. On that occasion, Stalin agreed to attack Japan within three months of Germany’s defeat: Russia had not declared war on Japan, as it was facing Hitler’s attack from the West. All the leaders were aware that the Manhattan Project was nearing success, although none of them mentioned it.

\vskip 0.4 cm

From Yalta onward, there was a dramatic acceleration of events: 

\begin{itemize}[itemsep=1pt,parsep=0pt]
		\item 
12 April: Roosevelt death.
    \item 
30 April: Hitler suicide.
     \item
8 - 9 May: Germany unconditional surrender.
    \item
11 June: The Franck Report from Los Alamos against using the Bomb on Japan, a move that would have triggered a nuclear arms race.
   \item 
12 July: Japan asks Russia to mediate for conditional surrender.
    \item 
16 July: Trinity (Alamogordo) test of the Bomb.
    \item 
17 July: Petition at Los Alamos requesting Truman not to use the Bomb against Japan. That petition is a reinstatement of the previous Franck report.
     \item
17 July: The winning nations meet again in Postdam. Truman “causally mentions Stalin that US has a new weapon of unusual destructive force”, but Stalin apparently do not show special interest. However, immediately after the meeting he tells his top advisors to speed up the atomic Bomb project that Russia had restarted towards the end of 1941.
       \item
26 July: Postdam Declaration: UK, USA and China indicate the terms of surrender to Japan, otherwise it would face “prompt and utter destruction”.
       \item
6 August: Atomic Bomb on Hiroshima.
       \item
8 August: URSS declares war to Japan, within three months of Germany surrender, as agreed in Yalta, and invades Manchuria.
       \item
9 August: Atomic Bomb on  Nagasaki.
        \item
15 August: Emperor Hirohito announces the Japan unconditional surrender.
\end{itemize}

It is certainly outside my competences to discuss the issue of the bombardment of Hiroshima and Nagasaki and the responsibility for that war crime, there are likely entire libraries devoted to the subject. However, as in the rest of this essay, I cannot help but note the somewhat surprising chronology of events. 

The allies had repeatedly asked Stalin to declare war on Japan, but he had refused, partly on the basis of an old non-aggression pact. At Yalta, however, he agreed to do so within three months of Germany’s defeat, and he did so exactly three months later, two days after Hiroshima. He played dumb. He knew that on the 12th of July Japan has asked Russia to mediate for a conditional surrender (the allies knew this as well, thanks to intercepted Japanese messages). In addition Stalin, informed by Truman, knew that America had the Bomb and had already decided to use it on Japan. Indeed the ultimatum to Japan was issued on 26 of July, just 10 days before the Bomb was dropped, when at least part of Japanese rulers was still hoping to secure a conditional surrender trough Russia. That hope was shattered not only by Hiroshima, but by the war declaration by Russia which followed immediately. Therefore Stalin obtained what he wanted (Manchuria) with minimum effort, leaving to the Anglo-American side the moral responsibility for the use of the Bomb against civil population. The moral responsibility of the “strategic bombing”, i.e. the bombardment of cities to weaken population resistance, was already felt by Churchill in England after the bombardment of Dresden (approximately 30000 civil causalities).

At least initially, Truman was unable to hide his pride for the success of the Hiroshima destruction: just sixteen hours after, on the 6th  of August, he (and the military) released the following public statement: ”...It is an atomic bomb. It is a harnessing of the basic power of Universe. The force of which the sun draws its power has been loosed against those who brought war to the Far East..” Hubris was taking over, the tone is triumphant, the Bomb looks like a divine punishment, and “..we are now prepared to obliterate.... every productive enterprise the Japanese have above the ground in any city." And concludes: “I (shall) make further recommendation to the Congress as to how atomic power can become a powerful and forceful influence towards the maintenance of world peace”. The last sentence seems to imply that the US have the power to maintain the peace thanks to the superiority gained through the Bomb. By now, we all know how false that is.

Even before the Trinity test, General Grooves, for his part, had commissioned Dr. H.D. Smyth the writing of a report entitled “The Atomic Energy for military purposes”, a booklet that come out in print on August 1945, the same month of Hiroshima. Grooves stated that purpose of the report was to inform the public about the story of the Bomb construction by the US, but in addition to providing transparency to the American people, the report had a clearly celebratory purpose, highlighting the achievements by the American side. In the preface Grooves indeed stated that "References to British and Canadian work are not intended to be complete since this is written from the point of view of the activities in this country".

Grooves was criticized for making public the process that led to the Bomb construction, but he replied that the American people needed to be informed after the secrecy that covered the project in war times. 
Furthermore, no sensible information had actually been disclosed: thanks to the compartmentalization of the project, Grooves had already ensured that no other nation - not even England - could build the Bomb. Only America possessed the ultimate weapon, every general’s dream. However, it was also clear that this advantage would not last forever: the Manhattan Project had to continue in order to maintain the lead. The new enemy, the communist USSR, was estimated to lag behind by about 10 years. However, as mentioned above, the Franck Report of June 1945 - drafted by a group of eminent physicists from the Manhattan Project following secret meetings - already foresaw the development of the nuclear arms race, as other countries, particularly the USSR, would develop their own nuclear weapons. But the military had already decided, and during the reorganization of the Manhattan Project in peacetime, many scientists left.

After the initial enthusiasm, Truman  tried to “justify” the destruction of Hiroshima and Nagasaki by arguing that it was the only way to stop the war and save the life of American soldiers facing the strenuous and suicidal (kamikaze) defense of Japaneses. That was only partly true. After the Trinity test, many scientists on the Manhattan Project again urged that the bomb they had helped build against Hitler not be used for hitting Japan, already on the verge of defeat. As an alternative they proposed to show Japan the terrible destruction power of the Bomb by an explosion in a desert area. The Truman advisors, however, were against it, because if the explosion had failed, the effect on Japan would have been counterproductive. In other words, the military wanted one "field test" of the Bomb - they eventually got two - before the war was over.

Another partly unforeseen problem was the deadly radioactive fallout, which would cause people die in excruciating pain months or even years thereafter. Although military tried to deny it, they were at least partially aware of, and after a  courageous report by The New Yorker journal,  America started to feel the blame. Truman and others still sought to defend the use of the Bomb by repeating that it had saved many young lives, both American and Japanese, but as it was clear since 1943-1944, the use of the Bomb was aimed at subduing URSS and the Communism. However, Stalin was not intimidated, and gave a dramatic boost to the URSS nuclear program, as predicted by the Franck report. Moreover, after the war Communism started to spread also in Asia, likely in response to centuries of European colonialism. With France and England weakened by the war, USA took over their place, as shown for instance in the Indochina wars, initially directed against France, and that ended in the US-Vietnam war once France was defeated at Dien Bien Phu. In addition, in 1946 civil war exploded between Mao's Communists and US-backed Ciang Kai-shek's Kuomintang party. Three years later the war ended with the victory of Communists in the mainland, with Kuomintang retreating to Taiwan with the protection of the US. Thus Communism took over in China, and immediately afterward a war started between Communist North Korea and US backed South Korea. Peace, therefore, was by no means guaranteed by the American monopoly on the atomic Bomb, and Communism was spreading everywhere. 

Amid this climate of political turmoil, it this rather obvious that the hostility to Communism in the American political and military circles, already present but concealed during the war, were increasingly growing and spreading among the public. Indeed, when in France Joliot built the first nuclear reactor, Zoé, on 15 December 1948, the Time US magazine published an article calling the reactor “A Communist Atomic Pile”, followed by New York Herald newspaper dubbing Zoé a “veritable threat”. An atomic reactor was not a nuclear weapon, but it was the first necessary step. Incidentally, Joliot, a convinced communist but also a pacifist, strongly believed that nuclear power should not be exploited to make a weapon. It is not surprising, therefore, that the news of the first nuclear explosion conducted by the Soviet Union on August 29, 1949, sparked fear and hostility toward Communism in the United States.

 The nuclear arms race had begun well before the US expected. With the usual sense of superiority, Americans could not believe that Soviet Union had scientists as clever as the US had. Therefore the URSS achievement was viewed as the result of espionage, which almost immediately led to witch hunt for spies within in the Manhattan project, and for all people that in the past had been sympathetic to Communism. Senator Joseph McCarthy exploited the general hysteria to gain political power. The second red-scare campaign had begun, communists at large were considered the enemy. Even inside Universities leftist or fully independent professors were declared unfit to teach. As a matter of fact, some of the scientists working at the Manhattan project were spying for URSS, even more than those that were eventually uncovered, but only for idealistic reasons. They wanted to fight Hitler and Nazifascism, which had emerged in Europe as a threat to Communism, and the United States was fighting fascism alongside the USSR. So why view the USSR and Communism as enemies, just as Nazism did? At the time, what Communism in the URSS had become under Stalin was largely unknown. According to present estimates, information transmitted to Soviet Union by the Manhattan scientists shortened the development of the Soviet Bomb by two years at most, and mostly helped to avoid dead-ends in the project. After all, Soviet Union already had top-level scientists, working with a dedication strengthened by the war-enhanced patriotism. 

Anyway, by 1948 the reality was that the US no longer had the monopoly of the ultimate weapon, that they could use or menace to use. This fact became evident during the Korea war: General McArthur, being unable to defect North Korea once and for all, proposed to use the Bomb against Korea and China. But considering that Soviet Union already had the Bomb, politicians dismissed McArthur and a truce was reached. 

After Hiroshima and Nagasaki, US conducted four more nuclear tests before the Soviet Union began its own. Once the nuclear arms race began, that number rose rapidly: over the next decade, US tests reached 188, an average of 18 per year. During the same period, the Soviet Union conducted 82 tests and the United Kingdom 21. Immediately after 1948, Truman ordered the development of the much more powerful Hydrogen, or thermonuclear Bomb, as from the project by Teller and Ulam. The US first test of the H-Bomb was in November 1952, but the Soviets followed suit only ten months later. So the nuclear testing race accelerated, but US faced an increasingly hostile public opinion, scared by the terrible consequences of the perhaps avoidable bombardment of Hiroshima and Nagasaki, and by the increasing power of the explosions and the consequent radioactive fallout. 

Consequently, in 1953 President Eisenhower launched the “Project Candor”, a public relation campaign aimed at gaining public support for the development of nuclear energy. One of the first initiatives was the “Atoms for Peace” campaign, aimed at building nuclear reactors for electricity generation under the supervision of the newly established Atomic Energy Commission (AEC). The “Atoms for Peace” program also aimed to promote the construction of nuclear reactors outside the United States, providing “friend” countries - such as Israel or Iran - with the necessary know-how, while concealing the fact that nuclear reactors are the first step toward building the Bomb. So in a sense the Atom for Peace campaign was to relieve the public scare of nuclear power, and at the same time to convince the American/Occidental people that more and more powerful Bombs had to be built against the evil Communist countries, Soviet Union in first place. Huge expenditures were therefore necessary to win the nuclear race and make America safer - although the risks connected with nuclear testing were contradicting that narrative. For instance, the “Castle Bravo” experiment for the second US thermonuclear Bomb, occurring in March 1954 at the Bikini Atoll, was about 2.5 times greater than scientist had predicted, and released a huge quantity of radioactive debris into the atmosphere, with consequent contamination of nearby atolls, US servicemen and unaware  Japanese fishermen.

In 1955 a group of 10 eminent scientists: Born, Bridgman, Einstein, Infeld, Joliot, Muller, Pauling, Powell, Rotblat, Russell and Yukawa published the so-called Russell-Einstein manifesto, pledging for the end of the nuclear race in the name of mankind, that would be destroyed by a nuclear war. Noteworthy is the presence of Joliot, who was later removed from the nuclear research program when France decided to enter the atomic Bomb race (the De Gaulle's "force de frappe"). Since Joliot never renounced his communist and pacifist beliefs, he was ostracized by his own colleagues, and faded into oblivion.  The Russell-Einstein manifesto gave rise to the so-called Pugwash community, a group of concerned scientists that starting on 1957 launched a series of Conferences on Science and World Affair, seeking for a world free of nuclear and other weapons of mass destruction. In the wake of World War two, this call for banning the most horrible mass destruction weapons had a considerable impact on public opinion. Distracting the public's attention through the Atoms for Peace campaign was not enough, people did not want the dangerous nuclear tests, that sometimes may go wrong, as in Castle Bravo, and that in any case release invisible radioactive debris that remain for years.  But testing the Bomb was considered essential by the military. It was perhaps under such conflicting pressures that in 1957 the military division of AEC launched the “Plowshare Project”  to study the technical and economic feasibility of using “peaceful nuclear explosives” for civilian and industrial applications. The name was inspired by the Bible verse (Isaiah 2:2-4): “ ... and they shall beat their swords into plowshares...”.

The idea was simple: for instance, ordinary explosives are used to build roads, mining etc., so why not to use the Bomb for the same aim, of course on a larger scale? For instance, digging harbors, shape mountains and so on. In the words of Edward Teller, a known outspoken advocate of the Plowshare Project: “We will change the earth's surface to suit us”, or “If your mountain is not in the right place, just drop us a card”. This is true hubris in the original Greek sense of the word: people thinking themselves more powerful than the Gods. Indeed, some of the projects put on paper but never realized had precisely this mindset. For instance: the world had just experienced the Suez Canal crisis, so why not build an alternative sea-level canal by using nuclear explosives? However, the area was deemed too unstable, and the project never progressed beyond the general concept. A more detailed plan was drawn up - complete with maps illustrating possible routes - for an alternative to the Panama Canal, which even included longer excavation routes, such as through Costa Rica and Nicaragua, or through Colombia. The project received \$17.5 millions for a detailed analysis, but it was later abandoned for technical reasons and for the opposition of the involved populations, that did not like the idea of being treated as guinea pigs for project they were not taking any advantage from. The same fate befell the Chariot project, an experiment aimed at digging a new harbor in Alaska, a sparsely populated US state. The military objective behind the project was the strategic location of the planned harbor, just 300 km from Siberia. The site was described as being far from any human settlements, but an indigenous Inuit community was located just 50 km away. Protests by the local population, combined with concerns about the fragility of the environment and the protection of the natural habitat, forced the AEC to abandon this project as well.

The first nuclear explosion under the Plowshare Program took place on December 10, 1961, three years after the project’s official launch. The delay was due to a moratorium on nuclear testing between the United States and the USSR, but as soon as the moratorium was lifted, the so-called Gnome Experiment took place. The AEC was eager to launch the Plowshare project: the Gnome Experiment consisted of a 3-kiloton (kT) underground nuclear explosion, with rather vague scientific objectives, such as the possibility of using heat to produce steam to be converted into electricity, or studying the feasibility of recovering radioisotopes for scientific and industrial applications. Actually the experiment was conceived as a kind of advertisement for the Plowshare Project, as well as a way to reassure the public: underground tests could be economically viable and, above all, were safe, since radioactive fallout could be contained and not released into the air. The AEC invited representatives from various media outlets and countries. From this perspective, the Gnome experiment was a complete failure: the bomb was buried 360 meters deep, but the ground above the blast site rose by nearly two meters. As newspapers reported, an “ominous-looking mushroom cloud emerged”, venting radiation into the atmosphere. So the advertised safe nuclear promise evaporated with the blast.

Despite the failure, the Plowshare program went on with two other, less advertised, underground explosions, always with vaguely stated scientific interest: one was Sedan in 1962, a thermonuclear explosion of 104 kT and extensive radioactive emission into the surrounding.  The generated crater is now a tourists' attraction. The other explosion was that of the little-known Sulky experiment in 1964. The stated scientific objective of these projects indicates that they were all funded by the AEC or by associated laboratories such as Livermore, whereas the Plowshare project was intended to generate a certain economic return by involving the private sector, as was already the case with the construction and operation of nuclear reactors. The next three experiments, Gasbuggy (1967) Rulison (1969) and Rio Blanco (1973) had the aim to extract natural gas, and received private funding.  All the bombs were detonated deep underground, and their nuclear yield was limited in order to contain the radioactivity. However, the resulting gas was contaminated - primarily with Tritium or Cesium 137 - and was therefore unusable. Thus, while over \$80 million was spent on these projects - partly funded by private industry but also by the government - no economic return was achieved; instead, there was only growing hostility from the public. Despite this striking series of failures, the Plowshare Program continued to propose other experiments, none of which were ever carried out, until its official end in 1977.

Just two years later, in 1979, the incident at the nuclear plant of Three Miles Island cast serious doubts about the safety of the controlled use of nuclear energy for civilian purposes as well. A recent book, "Atoms and Ashes", devoted to the history of the major nuclear accidents, lumps together Castle Bravo and Three Mile Island (and other well-known accidents such as Chernobyl and Fukushima), since the military and civilian uses of nuclear energy are closely intertwined, and in any case release lethal radioactive waste that can persist for thousands of years.

In any case, I believe that the push toward ending the nuclear arms race was driven not so much by public opinion, which had its importance in the West, but I don't know how much in the USSR, but rather by the realization that no one can win a nuclear war. Furthermore,
given the balance between the two major powers, it was necessary to devote significant effort and resources to the defense, in order to prevent the enemy from launching a devastating first strike that would preclude any response. It was therefore important to have nuclear weapons as close as possible to the enemy’s border, and to maintain
a permanent mobile fleet - aircraft or submarines - carrying nuclear weapons and able to strike back when necessary.

But if the entire responsibility for using the bomb is entrusted to relatively low-ranking military personnel, something can easily go wrong, as demonstrated by the Cuban Missile crisis (1962). On that occasion, the commander of a Soviet submarine, cut off from communications with high command and intercepted by an American warship enforcing the naval blockade around Cuba to prevent the USSR from installing there a nuclear missile base, decided to surrender rather than follow military protocol, thereby averting the outbreak of a nuclear war.
This near-miss incident prompted the two heads of state, Kennedy and Khrushchev, to begin communicating directly via a hotline and to reach a compromise on the deployment of nuclear weapons.
The USSR refrained from using Cuba as a nuclear missile base,
and at the same time the US withdrew its nuclear missile bases
from Turkey (the distance from Avana to Washington is approximately the same as that from Ankara to Moscow).

That direct contact marked the beginning of a thaw in relations between US and URSS, that in 1968 led, as a first step, to the Non-Proliferation Treaty: nations subscribing it would not try to build a Bomb, and at the same time nations that already possessed a nuclear arsenal would not increase its capabilities - after all, they had more than enough to destroy the whole world. About 25 year later, the subscription to the
Comprehensive Nuclear-Test-Ban Treaty was opened. Apart
from formal signatures, the Treaty led to a stop of tests by the main nuclear powers: URSS latest nuclear test took place in 1990, and the US one in 1992.

But there is no happy ending. The USSR no longer exists, nor does
Communism in Russia, but these days Western leaders
are pointing to the Russian regime as a threat, and directing their citizens toward war, with a renewed push for the civilian and military use of nuclear energy. It is therefore clear, in my view, that the issues are not political in nature (Communism versus Democracy) or religious (Judaism-Christianity versus Islam, as in the Middle East), but stem from the hubris and greed of a corrupt leadership that
needs an enemy to justify itself, as in Orwell's novel 1984, and believes
it can escape nuclear apocalypse, as in the ending of Kubrick's film Dr. Strangelove.

\section*{Disclaimer}
This is not an article written by an historian, with few exceptions
I did not have access nor I looked for the original documents. I am rather a "concerned scientist", a retired University Professor of Chemical Physics, interested in the history of science and scientists' biography. My curiosity was stimulated by
some articles appeared in Physics Today. I had to reconsider
the role of political figures during the historical period that saw
the development and exploitation of nuclear energy, which is why
I decided to share my findings in this essay.
  
\section*{Bibliography}

Many original documents can be found here:

https://atomicarchive.com/resources/documents/index.html

\vskip 0.2 cm

\noindent
Listed below are my main bibliographic sources, in alphabetical order:
\vskip 0.2 cm

\noindent
J. Albright and M.Kunstel, {\it Bombshell: The secret story of America's unknown atomic spy conspiracy}, Times Books-Random House, New York and Toronto, 1997.

\noindent
C. Day, {\it Cold War Science}, Physics Today, {\bf 69}, 8 (2016).

\noindent
M. Goldsmith, {\it Fr\'ed\'eric Joliot-Curie, a biography},
Lawrence and Wishart, London, 1976.

\noindent
A. Hagedorn, {\it Sleeper Agent, the atomic spy in America who got away}, Simon \& Schuster, New York, 2021.

\noindent
S. Nagamiya, {\it The atomic bomb: its history and the struggles
of scientists}, Eur.Phys.J. H {\bf 51}, 3 (2026).

\noindent
H. Pell, {\it "Peaceful" nuclear explosives ?}, Physics Today, {\bf 76},  34 (2023).

\noindent
S. Plokhi, {\it Atoms and Ashes: a global history of nuclear disasters},
W.W. Norton \& Company, New York, 2022.

\noindent
J. Rotblat, {\it Leaving the bomb project}, Bull. Atomic
Scientists, {\bf 41:7}, 16-19 (1985).

\noindent
H. D. Smyth, {\it Atomic energy for military purposes}, Princeton University Press, Princeton, 1945.

\noindent
W. Sweet, {\it France's Oppenheimer}, Physics Today, {\bf 78}, 46 (2025).

\noindent
N. Thorndike Greenspan, {\it Atomic Spy, The dark lives of Klaus Fuchs}, Penguin-Random House, New York, 2020.   

\end{document}